\documentclass[preprint,preprintnumbers,amsmath,amssymb]{revtex4}
\usepackage[abs]{overpic}
\usepackage{graphicx,subfigure}
\preprint{HUPD1304}
\begin{document}
\def\nn{\nonumber}
\def\beq{\begin{equation}}
\def\eeq{\end{equation}}
\def\bei{\begin{itemize}}
\def\eei{\end{itemize}}
\def\bea{\begin{eqnarray}}
\def\eea{\end{eqnarray}}
\def\ynu{y_{\nu}}
\def\nub{{\bar{\nu}}}
\def\Hp{{H^+}}
\def\ep{{e^+}}
\def\em{{e^-}}
\def\ydu{y_{\triangle}}
\def\ynut{{y_{\nu}}^T}
\def\ynuv{y_{\nu}\frac{v}{\sqrt{2}}}
\def\ynuvt{{\ynut}\frac{v}{\sqrt{2}}}
\def\s{\partial \hspace{-.47em}/}
\def\ad{\overleftrightarrow{\partial}}
\def\ss{s \hspace{-.47em}/}
\def\pp{p \hspace{-.47em}/}
\def\bos{\boldsymbol}
\title{
Charged Higgs and Neutral Higgs pair production \\
of weak gauge bosons fusion process 
in $e^+ e^-$ collision 
}
\author{Takuya  Morozumi, Kotaro Tamai}
\address{Graduate School of Science, Hiroshima University
Higashi-Hiroshima, 739-8526, Japan}
\begin{abstract}
Pair production of the neutral and charged Higgs boson is a unique
process which is a signature of two Higgs doublet model.
%. 
In this paper, we study the pair
production and their decays of the Higges in the neutrinophilic Higgs two 
doublet model. The pair production occurs through  $W$ and $Z$
gauge bosons fusion process.
In the neutrinophilic model, the vacuum expectation value (VEV) of the 
second Higgs doublet is small and is proportional to
the neutrino mass.  The smallness of VEV is associated with the approximate
global U(1) symmetry which is slightly broken.
Therefore, there is a suppression factor for the U(1) charge breaking process. 
The second Higgs doublet has U(1) charge
and its single production from the  gauge boson fusion 
violates the U(1) charge conservation and is suppressed strongly to occur. 
In contrast to the single production,
the pair production of the Higgses conserves U(1) charge
and the approximate symmetry does not forbid it.
To search for the pair productions in collider experiment,
we study the production cross section of 
a pair of the charged Higgs and neutral
Higgs bosons in $e^+ e^-$ collision with center of energy from $600$ (GeV) 
to $2000$ (GeV). The total cross section varies from  
$10^{-4}$(fb) to $10^{-3}$ (fb) for degenerate ($200$ GeV)
charged and neutral Higgses mass case.
The background process to the signal is gauge bosons pair
$W^+ + Z$ production and their decays.
We show the signal over background ratio
is about $2\% \sim 3\%$ by combining the cross section ratio with ratios of
branching fractions.
%\subjectindex{B40, B53}
\end{abstract}
\maketitle
\section{introduction}
While LHC already have started constraining many new physics models, 
there are a few aspects in the beyond standard models for which future $e^+ e^-$collider \cite{Behnke:2013xla, Baer:2013cma} 
can make unique search scenarios because of its clean environment.
 In this paper, we study the signature 
of the neutrinophilic two Higgs doublet model
\cite{Davidson:2009ha} in $e^+ e^-$
collision by focusing on the pair production and their decays of the charged Higgs and neutral 
Higgs bosons.

In the neutrinophilic model, a second Higgs doublet is introduced
and the neutrino masses are generated from the tiny  
VEV (vacuum expectation value) of the second Higgs doublet.
The new U(1) global symmetry is introduced. The second
Higgs doublet and right-handed neutrinos have the U(1) charge $+1$
and the other fields do not have that charge. 
The U(1) global symmetry is approximate and is broken explicitly
by the soft breaking bilinear term with respect to
the second Higgs doublet and to the standard 
model like Higgs doublet.
The tiny VEV of the second Higgs generated 
is proportional to the coefficient of the mass dimension two in the bilinear term.

In the model, any U(1) charge violating process is suppressed by 
the tiny VEV. It also implies that the probability amplitude
is suppressed and is proportional to neutrino mass.
An example of suppressed process is a single second Higgs production 
with gauge bosons fusion.  In contrast to the single second Higgs production, 
the pair production of the
second Higgs is the U(1) charge conserving process. Therefore,
they are not suppressed. The processes in this category are 
$Z^*(\gamma^*) \to H^+ H^-$, $W^+ + W^- \to  H^++ H^-$ and 
$W^+ + Z \to H^+ + X$ ($X=A, h$), where $H^+,A,$ and $h$ denote
the charged Higgs, CP odd Higgs and CP even Higgs in the second Higgs
doublet, respectively.

In LHC set up, the charged Higgs pair production
$p+p \to Z^*(\gamma^*) \to H^+ + H^-$ is studied in 
\cite{Davidson:2010sf}. 
In \cite{Figy:2008zd}, vector boson fusion into the light CP
even Higgs pairs is
studied at the LHC. In \cite{Dolan:2012ac}, di-Higgs production in
various scenario is discussed.
In \cite{Papaefstathiou:2012qe}, the standard model Higgs boson pair production is studied.
Also see reference \cite{Goertz:2013kp} for the ratio of the 
cross section of the single Higgs boson and the pair production cross section in the context of the standard model. 

In our work, in $e^+ e^-$ collision, 
the pair production of the charged Higgs($H^+$) and neutral Higgs
($X$) in the second Higgs doublet is studied.  
We derive the pair production cross section; $e^+ + e^- \to \overline{\nu_e} + e^-+H^+ + X$ ($X=A,h$).  
%U(1) charge is important for the decay process of .
%The unsuppressed decays of the neutral Higgses with U(1) charge
%are invisible as $X \to \nu \bar{\nu}$ and the final state includes
%the right-handed neutrino.

The paper is organized as follows. In section 1, we set up the Lagrangian
which is used in the calculation of charged Higgs and neutral Higgs production.
In section 2, we derive the expression of the cross sections
for the pair production from $e^+ + e^-$ collision. 
In section 3 , the cross sections including the various differential
cross sections are numerically computed and they are compared to the 
standard model background cross section.
In section 4, the decays of the charged Higgs and neutral Higgs are discussed
and the dependence on the charged lepton flavor in the final state is studied.
Section 5 is devoted to the summary. 
\section{Two Higgs doublet model with softly broken global symmetry}
In this section, we present the Lagrangian to set up the notation and
also to display the interaction terms which are relevant to the
calculation in later sections.
The Higgs potential is given by \cite{Davidson:2009ha},
\bea
V_{\rm tree}&=&\sum_{i=1,2}\left(m_{ii}^2 \Phi_i^\dagger \Phi_i+\frac{\lambda_i}{2}
(\Phi_i^\dagger \Phi_i)^2 \right)-(m_{12}^2 \Phi_1^\dagger \Phi_2+h.c.) \nn \\ 
&+&\lambda_3 (\Phi_1^\dagger \Phi_1)(\Phi_2^\dagger \Phi_2) + \lambda_4 |\Phi_1^\dagger \Phi_2|^2.
\eea 
Two Higgs doublets in the unitary gauge are parameterized as
\cite{Morozumi:2011zu},
\bea
\Phi_1&=&\begin{pmatrix} 0 \\ \frac{v \cos \beta}{\sqrt{2}} \end{pmatrix}+
\begin{pmatrix} -\sin \beta H^+ \\
                 \frac{\sin \gamma h + \cos \gamma H-i \sin \beta A}{\sqrt{2}} 
\end{pmatrix}, \nn \\
\Phi_2&=&\begin{pmatrix} 0 \\ \frac{v \sin \beta}{\sqrt{2}} \end{pmatrix}+
\begin{pmatrix} \cos \beta H^+ \\
                 \frac{\cos \gamma h - \sin \gamma H+i \cos \beta A}{\sqrt{2}} 
\end{pmatrix}.
\eea
The new U(1) charge for $\Phi_1$ ($\Phi_2$) is $0(+1)$.
The term proportional to $m_{12}$ is U(1) breaking term.
H and h denote CP even Higgses. $A$ denotes a CP odd Higgs.
In our notation, H is close to the standard model like Higgs,
a different notation from \cite{Davidson:2009ha}. In most of the present paper,
we follow the notation of 
\cite{Morozumi:2011zu}.
$\tan \beta$ is the ratio of two VEVs and is given approximately as 
\cite{Davidson:2009ha},
\bea
\tan \beta=\frac{m_{12}^2}{m_A^2}.
\eea
$v^2$ is the squared sum of two VEVs. 
$\gamma$ is a mixing angle of CP even Higgses given by \cite{Morozumi:2011zu},
\bea
\tan 2\gamma=-\frac{-4 m_{12}^2+ 2 \sin 2\beta (\lambda_3+\lambda_4)v^2}
{(3(-\lambda_1 \cos \beta^2+\lambda_2 \sin^2 \beta)+\cos 2 \beta 
(\lambda_3+\lambda_4))v^2-2(m_{11}^2-m_{22}^2)}.
\eea
Then one can write the covariant derivative terms for the two doublets,
which includes the electroweak interactions of the Higges 
with gauge bosons,
\bea
&&\sum_{i=1,2} D_\mu \Phi_i^\dagger D^\mu \Phi_i \ni
gM_W (W_\mu^+ W^{\mu -}+\frac{1}{2 c_W^2} Z^\mu Z_\mu)
\left(\sin(\beta+\gamma) h + \cos(\beta +\gamma) H \right)\nn \\
&+& \frac{g^2}{2} s_W (A_\mu -t_W Z_\mu)[(H^+ W^{\mu -}+H^- W^{\mu +})
(h \cos(\beta+\gamma)-H \sin(\beta+\gamma))\nn \\
&-&i(H^+ W^{\mu -}-H^- W^{\mu +})
A] \nn \\
&+& i \frac{g \cos 2 \theta_W }{2 \cos \theta_W} Z_\mu (\partial^\mu H^- H^+-
\partial^\mu H^+ H^-) \nn \\
&+& \frac{g \cos(\beta+\gamma)}{2 \cos \theta_W}
(\partial_\mu h A- \partial_\mu A h) Z^\mu \nn \\
&+&\Bigl{\{} i \frac{g}{2} \cos(\beta+\gamma) W^{+ \mu}(h \partial_\mu H^-- \partial_\mu h H^-)
+\frac{g}{2} 
W^{+\mu}(H^-\partial_\mu A- A \partial_\mu H^-) + h.c. \Bigr{\}}.
\eea
One notes that a single CP even Higgs boson ($h$ or $H$),
could be produced by the gauge boson fusion process 
$W^+ + W^-(Z + Z) \to h$ or $H$. 
There is no single CP odd Higgs $A$ production from
gauge boson fusion. Absence of the term like $A W_\mu^+ W^{- \mu}$ is
due to CP symmetry. We also note that CP even Higgs $h$ is mostly 
the real part of the down component of the
second Higgs $\Phi_2$. Its coupling to gauge boson pair operators
$W^{+\mu} W^-_\mu$ and $Z^\mu Z_\mu$ is suppressed as 
$\sin(\beta+\gamma)$.
Since $\sin \beta$ and $\sin \gamma$ are suppressed to be zero
in the vanishing limit of U(1) breaking term $m_{12}$, the
gauge boson fusion to $h$ is forbidden in the limit.  
As for the decays of charged Higgs and neutral Higgs, the Yukawa coupling
to right handed neutrino is important.
Assigning the U(1) charge $+1$ to right handed neutrino \cite{Davidson:2009ha},
it is written in terms
of mass eigen states as,
\bea
{\mathcal L}_Y&=&-y_{\nu ij} \overline{\psi_i} \tilde{\Phi}_2 \nu^0_{Rj} \nn \\
          &\ni&-\overline{\nu_i}(\frac{m_{\nu i}}{v})
 \nu_i \frac{\cos \gamma h-\sin \gamma H}{\sin \beta}+ i 
\overline{\nu_i}(\frac{m_{\nu i}}{v}) \gamma_5 \nu_i
\cot \beta A \nn \\
&+& \sqrt{2} \cot \beta 
\overline{l_i} V_{ij} (\frac{m_{\nu j}}{v})\nu_{Rj} H^- + h.c.,
\label{eq:Yukawa}
\eea
where $m_\nu$ denote neutrino masses and $V$ denotes Maki Nakagawa Sakata 
(MNS) matrix.
\section{Cross section for
$e^+ + e^- \rightarrow \bar{\nu}+ e^- + W^{+ \ast}+ Z^\ast 
\rightarrow \bar{\nu}+ e^-+ H^+ + A $}
In this section, we present the formulae for the cross section of 
$e^+ + e^- \rightarrow \bar{\nu}+ e^- +W^{+ \ast} + Z^\ast 
\rightarrow \bar{\nu} +e^- +H^+ + A $. (See Fig.1.)
\begin{figure}
\begin{center}
\includegraphics[width=10cm]{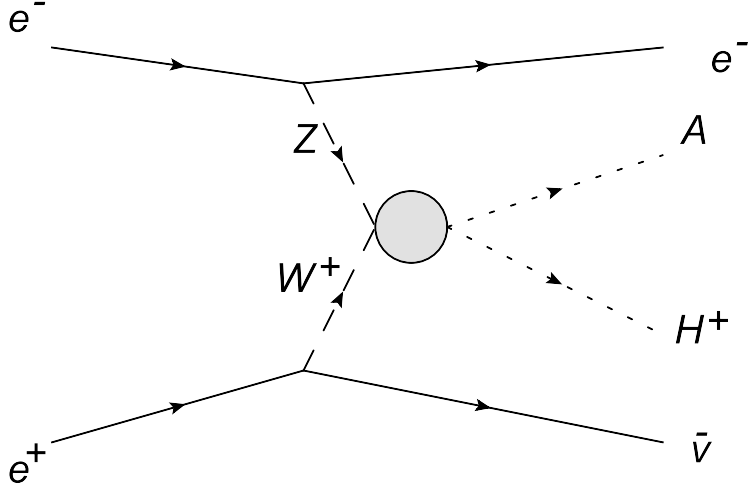}
\caption{Feynman diagram of charged Higgs $H^+$ and CP odd Higgs
$A$ production in $e^+ e^-$ collision. The production
occurs through $W^+$ and $Z$ fusion which is shown in
the circle.}
\end{center}
\end{figure}
We define,
\bea
\sigma_{H^+ X}&\equiv&\sigma(e^+ + e^- \rightarrow 
\overline{\nu_e} + e^-+H^+ + X)
; X= A, h.
\eea
We write the cross section for $\Hp A$ production, 
\bea
\sigma_{H^+A}&=&\frac{1}{2s_{\ep \em}}\int \frac{d^3 q_A}{(2 \pi)^3 2 E_A}
\frac{d^3 q_{H^+}}{(2 \pi)^3(2 E_H^+)} \frac{d^3 q_e}{(2 \pi)^3 2 E_e(q_e)}
\frac{d^3 q_\nu}{2 E_\nub} \nn \\
&\times& \frac{1}{4} \sum_{\rm spin} |M|^2 (2 \pi)  
\delta^4(p_\ep + p_{e} 
-q_{H^+}-q_A-q_e-q_\nub).
\eea
$s_{\ep e^-}$ is the center of mass (cm) energy of 
$\ep$ and $e^-$ collision. $p_\ep$ and $p_{e}$
denote the momentum of positron and electron of the initial
state. $q_e$ $q_{H^+}$,$q_A$ and $q_{\bar{\nu}}$ are momentum of the 
final states; i.e., electron, charged Higgs,
neutral Higgs and anti-neutrino respectively.
The transition amplitude $M$ is given by, 
\bea
M=-T_{A \mu \nu} \frac{1}{(p_Z^2-M_Z^2)(p_W^2-M_W^2)}
\frac{g^2}{2 \sqrt{2} c_W}\overline{u(q_e)}
\gamma^\nu (L +2 s_W^2) u(p_e) \overline{v_{\ep}(p_\ep)}\gamma^\mu L 
v_{\nub}(q_\nub).
\eea
where $p_Z=p_e-q_e$ and $p_W=q_{H^+}+q_A-p_Z$.
$L$ denotes the chiral projection $L=\frac{1-\gamma_5}{2}$. 
$s_W (c_W)$ denotes sine (cosine) of the Weinberg angle.
$T_{A \mu \nu}$ denotes the off shell amplitude for 
$W_\mu^{+ *} + Z_\nu^{*} \rightarrow A+ H^+$ production.
It corresponds to the circle in Fig.1 and the Feynman diagrams
which contribute to $T^{A}_{\mu \nu}$ are shown in 
Fig.2 $\sim$ Fig.5.
\begin{figure}[htb]
\begin{center}
\includegraphics[scale=0.6]{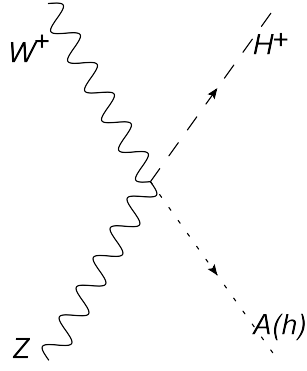}
\caption{Contact interaction}
\end{center}
\label{fig:contact}
\end{figure}
\begin{figure}[htb]
\begin{tabular}{ccc}
\begin{minipage}{0.28\hsize}
\begin{center}
\includegraphics[scale=0.5]{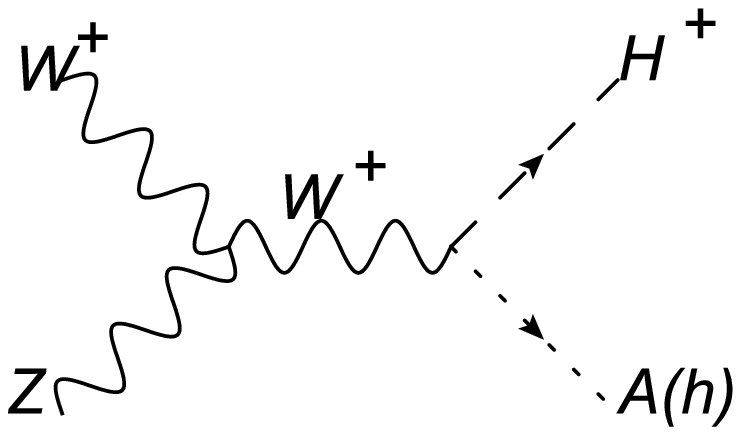}
\caption{S channel W exchange}
\end{center}
\end{minipage}
\begin{minipage}{0.28\hsize}
\begin{center}
\includegraphics[scale=0.5]{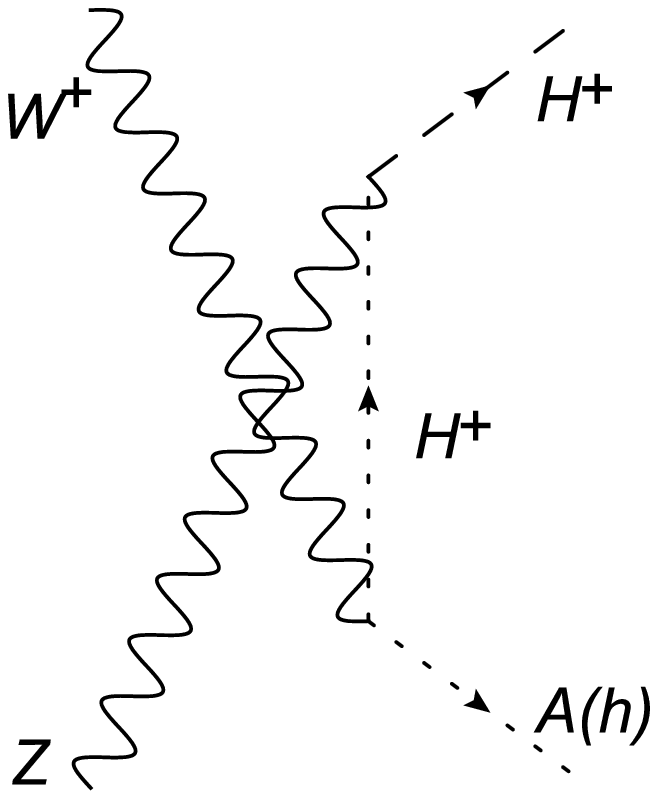}
\caption{U channel}
\end{center}
\end{minipage}
\begin{minipage}{0.28\hsize}
\begin{center}
\includegraphics[scale=0.5]{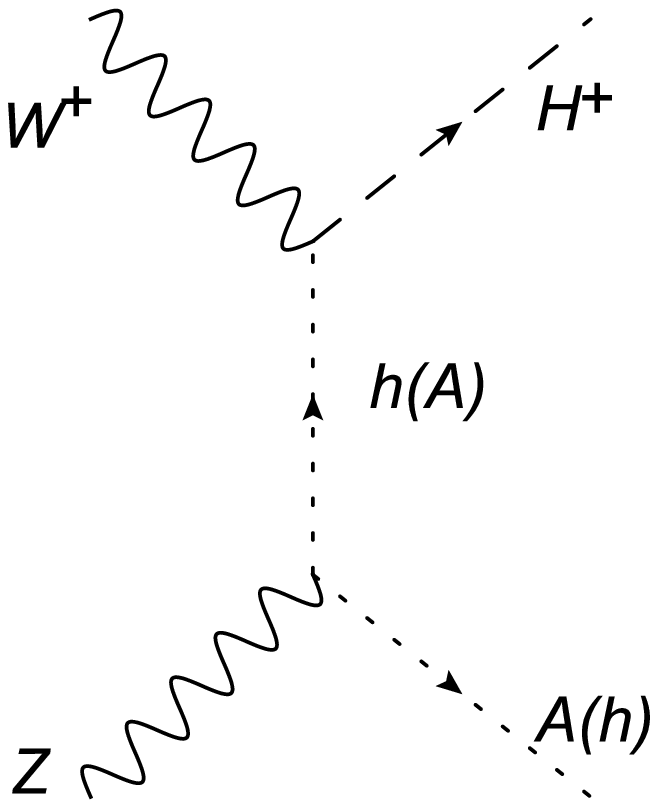}
\caption{T channel}
\end{center}
\end{minipage}
\end{tabular}
\label{fig:tchan}
\end{figure}
The second rank tensor $T_{A \mu \nu}$ is given as,
(On-shell case is shown in \cite{Morozumi:2012sg}.)
\bea
T_{\mu \nu}=i T_{A \mu \nu}&=&
\frac{g^2}{2\cos \theta_W}
\left(a_A g_{\mu \nu}+ d_A q_{A \nu} q_{H^+ \mu}
+b_A  q_{H^+ \nu} q_{A \mu}\right).
\label{eq:Aamp}
\eea 
where we introduce the real amplitude $T^\ast_{\mu \nu}=T_{\mu \nu}$.
$a_A, b_A$ and $d_A$ in Eq.(\ref{eq:Aamp}) are given as,
\bea
a_A&=&s_W^2+\frac{p_Z^2-p_W^2}{M_Z^2}
\frac{M_A^2-M_{H^+}^2-M_W^2}{s_{\Hp A}-M_W^2}+
c_W^2 \frac{t_A-u_A+p_Z^2-p_W^2}{s_{\Hp A}-M_W^2}, \nn \\
b_A&=&-\frac{2\cos 2 \theta_W}{u_A-M_{H^+}^2}- 
\frac{2(\cos 2 \theta_W+1)}{s_{\Hp A}-M_W^2},\nn \\
d_A&=&\frac{2 \cos^2(\beta+\gamma)}{t_A-M_h^2}+
\frac{2(\cos 2 \theta_W+1)}{s_{\Hp A}-M_W^2},
\eea
with $t_A=(q_{H^+}-p_W)^2, u_A=(p_W-q_A)^2$ and $s_{\Hp A}=(q_\Hp+q_A)^2$.
The spin averaged amplitude squared is given as,
\bea
\frac{1}{4} \sum_{\rm spin} |M|^2=
\frac{g^4}{32 c_W^2}\frac{1}{|(p_Z^2-M_Z^2)(p_W^2-M_W^2)|^2} 
T_{\mu \nu} T^\ast_{\rho \sigma}  L_{ee}^{\nu \sigma}
L_{e^+ \nub}^{\mu \rho},
\eea
where $L^{\nu \rho}_{ee}$ is a leptonic tensor of the neutral current 
and $L^{\mu \sigma}_{e^+ \nub}$ is the one of the charged current.
They are written in terms of the symmetric part $S$ and the anti-symmetric
part $A$. 
\bea
L_{ee}^{\nu \sigma}&=&S_{ee}^{\nu \sigma}+ i A_{ee}^{\nu \sigma}, \nn \\
S_{ee}^{\nu \sigma}&=&(2+ 8 s_W^2+ 16 s_W^4)
(p_e^\nu q_e^\sigma- g^{\nu \sigma} p_e \cdot q_e +  p_e^\sigma q_e^\nu),\nn \\
A_{ee}^{\nu \sigma}
&=&(2+8 s_W^2) \epsilon^{\nu \alpha \sigma \beta} p_{e \alpha}
q_{e \beta}, \\
L_{e^+ \nub}^{\mu \rho}&=&S_{e^+ \nub}^{\mu \rho}+ i 
A_{e^+ \nub}^{\mu \rho}, \nn \\
S_{e^+ \nub}^{\mu \rho}&=&2 (q^\mu_\nub p^\rho_{e^+}-g^{\mu \rho} q_\nub \cdot p_{e^+}
+q^\rho_\nub p^\mu_{e^+}), \nn \\
A_{e^+ \nub}^{\mu \rho} &=&2 \epsilon^{\mu \alpha \rho \beta}
q_{\nub \alpha} p_{e^+ \beta}.
\eea
We define the transpose matrix as $T^t_{\mu \nu}=T_{\nu \mu}$. 
In terms of them, one can write the differential cross section as,
\bea
d \sigma_{H^+ A}&=& \frac{g^4}{64 c_W^2 s_{\ep \em}}\frac{1}{4096 \pi^8}
\Biggl{|}\frac{1}{((p_e-q_e)^2-M_Z^2)((p_e^+-q_\nub)^2-M_W^2)}
\Biggr{|}^2 \nn \\
&&(T_{\mu \nu} S_{ee}^{\nu \sigma} T^t_{\sigma \rho} S_{e^+ \nub}^{\rho \mu}
+T_{\mu \nu} A_{ee}^{\nu \sigma} T^t_{\sigma \rho} A_{e^+ \nub}^{\rho \mu})
d^{12} Ph,
\eea
where $d^{n}Ph$ denotes $n$ dimensional phase space integral.
For $n=12$, it is defined as, 
\bea
d^{12} Ph=\frac{d^3 q_A d^3 q_\Hp d^3 q_e d^3 q_\nub}{E_A E_\Hp E_e E_\nub} 
\delta^4(p_\ep+ p_e-q_e-q_\nub-q_\Hp-q_A). 
\eea
In center of mass frame of $ e^+ e^-$ collision, the amplitude is independent
of the rotation around the beam axis. One can also set the direction of
the $e^+$ beam to z direction and the momentum of electron in the final states
in $y z$ plane. 
Therefore after one integrates the azimuthal angle and 
the anti-neutrino momentum,  one obtains $d^8 Ph$ as, 
\bea
d^{8}Ph&=& 2 \pi d \cos \theta_e d \cos \theta_{eH} d \phi_{eH}
d \cos \theta_{eHA} d \phi_{eHA}\nn \\
&& \frac{q_e^2 dq_e}{E_e} \frac{q_{\Hp}^2 d q_{\Hp}}{E_\Hp} \frac{q_{A}^2 d q_{A}}{E_A} 
\delta(\sqrt{s}-E_\Hp-E_A-E_e-E_\nub).
\eea
The momentum of electron $q_e$ in final states is specified by a polar angle 
$(\theta_e)$ in the orthgonal frame in which positron momentum is chosen
as z axis. 
\bea
{\bf p_\ep}&=&\frac{\sqrt{s_{\ep e^-}}}{2} {\bf{e_3}}, \quad {\bf p_e}=
-\frac{\sqrt{s_{\ep e^-}}}{2} {\bf e_3}, \nn \\
{\bf q_e}&=&|{\bf q_e}|(
\sin \theta_e {\bf e_2} + \cos \theta_e {\bf e_3}), \nn \\
{\bf e_1}&=& {\bf e_2} \times {\bf e_3}.
\eea
One can define a new orthogonal coordinate spanned by the basis vectors
${\bf e^\prime_i} (i=1 \sim 3)$.
\bea
{\bf e_3^\prime}&=&\frac{{\bf q_e}}{|\bf q_e|}=\sin \theta_e {\bf e_2} + \cos \theta_e {\bf e_3}, \nn \\
{\bf e_2^\prime}&=&-\sin \theta_e {\bf e_3} + \cos \theta_e {\bf e_2}
,\nn \\
{\bf e_1^\prime}&=&{\bf e_1}.
\eea
$\theta_{eH}$ and $\phi_{eH}$ denote the momentum direction of 
the charged Higgs
relative to that of the electron in the final state.
\bea
{\bf q_\Hp}=|{\bf q_{\Hp}}|(\sin \theta_{eH} \cos \phi_{eH} {\bf e_1^\prime}
+ \sin \theta_{eH} \sin \phi_{eH} {\bf e_2^\prime}+ \cos \theta_{eH} \bf{e_3^\prime}).
\eea
Finally $(\theta_{eHA}, \phi_{eHA})$ denote the direction of momentum for 
the neutral Higgs $A$. $\theta_{eHA}$ is a polar angle 
measured from the direction ${\bf q_e+q_\Hp}$. 
\bea
\bf{q_A}&=&|q_A|(\sin \theta_{eHA} \cos \phi_{eHA} {\bf e_1^{\prime \prime}}
+ \sin \theta_{eHA} \sin \phi_{eHA} {\bf e_2^{\prime \prime}} + \cos \theta_{eHA} {\bf e_3^{\prime \prime}}), \\
{\bf e_3^{\prime \prime}}&=&
\frac{{\bf q_e}+{\bf q_{\Hp}}}{|{\bf q_e}+{\bf q_{\Hp}}|},
{\bf e_1^{\prime \prime}}=\frac{{\bf q_e \times q_\Hp}}{|{\bf q_e \times q_\Hp}|} , 
{\bf e_2^{\prime \prime}}= {\bf e_3^{\prime \prime}} \times 
{\bf e_1^{\prime \prime}}.
\eea
In terms of the angles defined, the phase space integration is written,
\bea
d^8Ph &=& 2 \pi d \cos \theta_e d \cos \theta_{eH} d \phi_{eH}
d \cos \theta_{eHA} d \phi_{eHA}\nn \\
&& \frac{q_e^2 dq_e}{E_e} \frac{q_{\Hp}^2 d q_{\Hp}}{E_\Hp} \frac{q_{A}^2 d q_{A}}{E_A E_\nub} 
\delta(\sqrt{s}-E_\Hp-E_A-E_q-E_\nub) \nn \\
E_\nub&=& \sqrt{|{\bf q_e+q_{\Hp}}|^2+q_A^2+ 2 \cos \theta_{eHA} q_A |{\bf q_e+q_{\Hp}}|},
\eea
where we denote $q_A=|{\bf q_A}|, q_\Hp=|{\bf q_\Hp}|$ and $q_e=|{\bf q_e}|$.
The integration over the variable $\cos \theta_{eHA}$ 
is carried out and we obtain,
\bea
d^7 Ph&=& 2 \pi d \cos \theta_e d \cos \theta_{eH} d \phi_{eH} d \phi_{eHA}
\frac{q_A}{E_A} 
dq_A \frac{q_\Hp^2}{E_\Hp} dq_\Hp q_e dq_e \frac{1}{|{\bf q_e+q_\Hp}|} \nn \\
&\times& \theta(E_\nub^0-||{\bf q_\Hp}+{\bf q_e}|- q_A||) 
\theta(|{\bf q_e}+{\bf q_\Hp}|+q_A-E_\nub^0),
\label{eq:d7ph}
\eea
where,
\bea
E_\nub^0=\sqrt{s_{\ep \em}}-E_e-E_A-E_\Hp.
\eea
The step functions in Eq.(\ref{eq:d7ph}) imply the phase space boundaries.
Using Eq.(\ref{eq:d7ph}), the differential cross section is,
\bea
&& \frac{d^7 \sigma_{\Hp A}}{d q_e dq_\Hp dq_A d \cos_e d \cos_{eH} d \phi_e 
d \phi_{eHA}}\\
&& = \frac{g^4}{32 c_W^2 s}\frac{1}{4096 \pi^7}
\Biggl{|}\frac{1}{((p_e-q_e)^2-M_Z^2)((p_e^+-q_\nub)^2-M_W^2)}
\Biggr{|}^2 \nn \\
&&(T_{\mu \nu} S_{ee}^{\nu \sigma} T^t_{\sigma \rho} S_{e^+ \nub}^{\rho \mu}
+T_{\mu \nu} A_{ee}^{\nu \sigma} T^t_{\sigma \rho} A_{e^+ \nub}^{\rho \mu})
\frac{q_A}{E_A} 
\frac{q_\Hp^2}{E_\Hp} q_e \frac{1}{|{\bf q_e+q_\Hp}|}\nn \\
&&\theta(E_\nub^0-||{\bf q_\Hp}+{\bf q_e}|- q_A||) 
\theta(|{\bf q_e}+{\bf q_\Hp}|+q_A-E_\nub^0).
\eea
We carry out the rest of integration numerically.
\section{Numerical results}
In this section, we present the numerical results for the cross
sections. 
We have carried out the phase space integrations by using the montecarlo
program, bases \cite{Kawabata:1995th}.
We have studied the three sets of the charged Higgs
and neutral Higgs masses. 
\bea
(m_\Hp, m_A)=(300,200), (200,300), (200, 200) (\rm{GeV}). 
\eea
As shown in \cite{Morozumi:2011zu}, for those
input values of charged Higgs and neutral Higgs masses,
the radiative corrections to the VEVs, $\beta$ and $v$
are within $10 \%$. 

\begin{figure}
\begin{center}
\includegraphics[width=10cm]{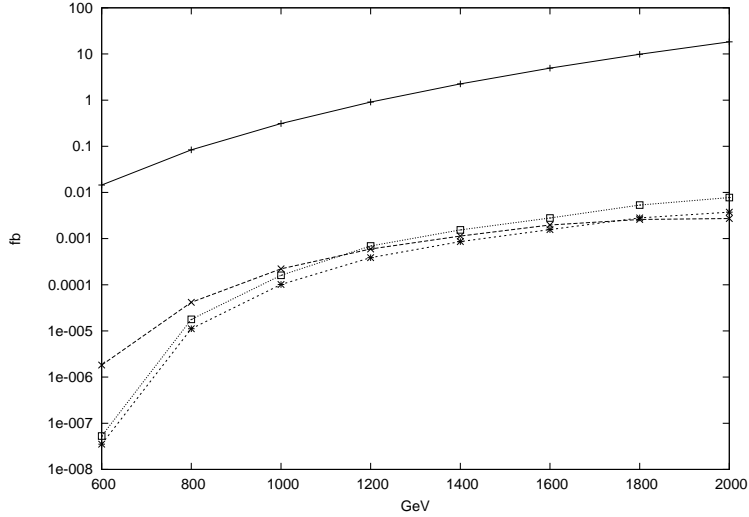}
\caption{The gauge boson pair production cross section ($\sigma_{WZ}$) for
$e^+ + e^- \rightarrow W^+ + Z + \overline{\nu_e}+ e^-$ 
(solid line) and the Higgs pair production cross sections ($\sigma_{\Hp A}$)
for $e^+ + e^- \rightarrow H^+ + A + \overline{\nu_e}+ e^- $.
The horizontal axis denotes center of mass energy; $\sqrt{s_{e^+ e^-}}$(GeV)
of $e^+ e^-$ collision. 
The long dashed line with the cross symbol $\times$ corresponds to the case 
$(m_{H^+},m_A)=(200,200)$(GeV). The dotted
line with the boxes $\Box$ corresponds to $(m_{H^+},m_A)=(300,200)$(GeV)
and the short dashed line with asterisks $*$ corresponds to
$(m_{H^+},m_A)=(200,300)$(GeV). 
}
\end{center}
\label{fig:cross}
\end{figure}

\begin{figure}
\includegraphics[width=15cm]{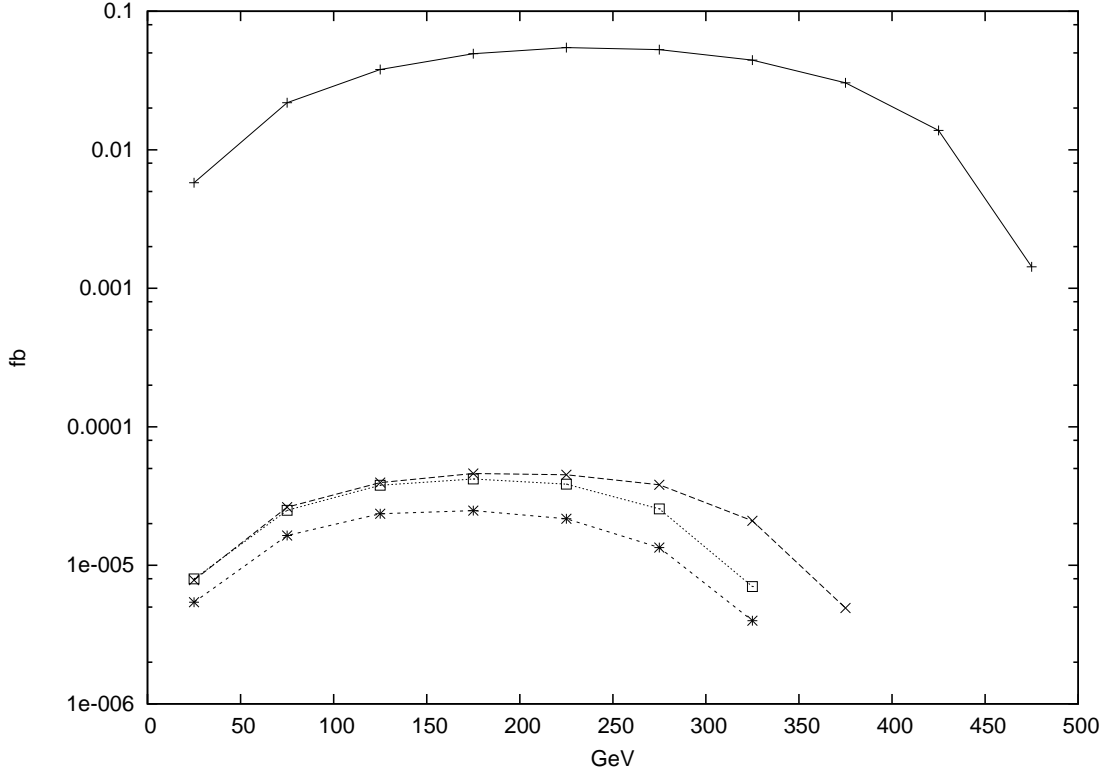}
\caption{The differential cross sections $\Delta \sigma_{1\Hp A}$
and $\Delta \sigma_{1 WZ}$ as functions of the momentum $q_e$(GeV) 
for the final 
state electron. We have chosen the width of each bin as $\Delta q_e=50$(GeV).
The solid line marked with the plus sign $+$ corresponds to 
$e^+ + e^- \rightarrow W^+ + Z + \overline{\nu_e}+ e^-$.
The other lines denote the three cases for  
$e^+ + e^- \rightarrow H^+ + A + \overline{\nu_e}+ e^- $.
The long dashed line marked with cross symbol $\times$ corresponds to the case 
$(m_{H^+},m_A)=(200,200)$(GeV). The dotted
line marked with the boxes; $\Box$ 
corresponds to $(m_{H^+},m_A)=(300,200)$(GeV)
and the short dashed line marked by asterisks $*$ corresponds to
$(m_{H^+},m_A)=(200,300)$(GeV). The center of mass energy is $1000$ (GeV).  
}
\label{fig:sigma1}
\end{figure}

We have shown the total cross sections $\sigma_{\Hp A}$ 
with respect to the center of mass energy $(\sqrt{s_{\ep \em}})$ of
$e^+ e^-$ collision
in Fig.\ref{fig:cross}.
Then 
we have plotted the following one
 dimensional differential cross sections;
Fig.\ref{fig:sigma1} $ \sim $ Fig.\ref{fig:sigma5}.
\bea
\Delta \sigma_{1\Hp A}(q_e)&=& \int_{q_e-\frac{\Delta q_e}{2}}^{q_e+\frac{\Delta q_e}{2}}\frac{d \sigma_{\Hp A}}{d q_e}  d q_e, \quad \Delta q_e=50 (\mbox{GeV}), 
\label{eq:difc1}
\\
\Delta \sigma_{2 \Hp A}(q_{\Hp})&=& \int_{q_\Hp-\frac{\Delta q_{\Hp}}{2}}^{q_\Hp+\frac{\Delta q_{\Hp}}{2}}
\frac{d \sigma_{\Hp A}}{d q_\Hp}  d q_\Hp, \quad \Delta q_\Hp=50 (\mbox{GeV}), 
\label{eq:difc2}
\\
\Delta \sigma_{3 \Hp A}(\cos \theta_e)&=& \int_{\cos\theta_e-\frac{\Delta \cos \theta_e}{2}}^{\cos\theta_e+\frac{\Delta \cos \theta_e}{2}}
\frac{d \sigma_{\Hp A}}{d \cos \theta_e} d \cos \theta_e, 
\quad \Delta \cos \theta_e=0.2,
\label{eq:difc3}
\\
\Delta \sigma_{4 \Hp A}(\cos \theta_{eH}) &=& 
\int_{\cos\theta_{eH}-\frac{\Delta \cos \theta_{eH}}{2}}^{\cos\theta_{eH}+\frac{\Delta \cos \theta_{eH}}{2}}
\frac{d \sigma_{\Hp A}}{d \cos \theta_{eH}} d \cos \theta_{eH},\quad \Delta \cos \theta_{eH}=0.2,
\label{eq:difc4}
\\
\Delta \sigma_{5 \Hp A}
(\phi_{eH})&=& \int_{\phi_{eH}-\frac{\Delta \phi_{eH}}{2}}^
{\phi_{eH}+\frac{\Delta \phi_{eH}}{2}}
\frac{d \sigma_{\Hp A}}{d \phi_{eH}} d \phi_{eH}. \quad \Delta \phi_{eH}=\frac{\pi}{5}.
\label{eq:difc5}
\eea
For comparison, we have also computed 
gauge boson production cross section. We have used the formulae
in \cite{Bahnik:1997ka} 
for $W+Z \to W+Z$ scattering amplitude. 

\bea
\sigma_{WZ}&\equiv&\sigma_{SM}(e^+ + e^- \rightarrow 
\overline{\nu_e} + e^-+W^+ + Z).
\eea
We have plotted 
$\sigma_{WZ}$ in Fig.\ref{fig:cross} as well as 
the differential ones; $\Delta\sigma_{i WZ }
(i=1 \sim 5)$ for the weak gauge boson pair ($W^+$ and $Z$)production
in the standard model. See Fig.\ref{fig:sigma1} $\sim$ Fig.
\ref{fig:sigma5}.
It can be a background process to the Higgs pair production.
Explicitly, we write the differential cross section 
$ \Delta \sigma_{i WZ}$ ($i=1 \sim 5$), which are defined 
analogous to the ones defined for the case of 
Higgs production in Eq.(\ref{eq:difc1}) $\sim$
Eq.(\ref{eq:difc5}).
\bea
\Delta \sigma_{1WZ}(q_e)&=& \int_{q_e-\frac{\Delta q_e}{2}}^{q_e+\frac{\Delta q_e}{2}}\frac{d \sigma_{WZ}}{d q_e}  d q_e, \quad \Delta q_e=50 (\mbox{GeV}), \\
\Delta \sigma_{2WZ}(q_{W})&=& \int_{q_W-\frac{\Delta q_{W}}{2}}^{q_W+\frac{\Delta q_{W}}{2}}
\frac{d \sigma_{WZ}}{d q_W}  d q_W, \quad \Delta q_W=50 (\mbox{GeV}), \\
\Delta \sigma_{3WZ}(\cos \theta_e)&=& \int_{\cos\theta_e-\frac{\Delta \cos \theta_e}{2}}^{\cos\theta_e+\frac{\Delta \cos \theta_e}{2}}
\frac{d \sigma_{WZ}}{d \cos \theta_e} d \cos \theta_e, 
\quad \Delta \cos \theta_e=0.2,
\\
\Delta \sigma_{4WZ}(\cos \theta_{eW}) &=& \int_{\cos\theta_{eW}-\frac{\Delta \cos \theta_{eW}}{2}}^{\cos\theta_{eW}+\frac{\Delta \cos \theta_{eW}}{2}}
\frac{d \sigma_{WZ}}{d \cos \theta_{eW}} d \cos \theta_{eW},\quad \Delta \cos \theta_{eW}=0.2,
\\
\Delta \sigma_{5WZ}(\phi_{eW})&=& \int_{\phi_{eW}-\frac{\Delta \phi_{eW}}{2}}^
{\phi_{eW}+\frac{\Delta \phi_{eW}}{2}}
\frac{d \sigma_{WZ}}{d \phi_{eW}} d \phi_{eW}, \quad \Delta \phi_{eW}=\frac{\pi}{5}.
\eea
\begin{figure}
\includegraphics[width=15cm]{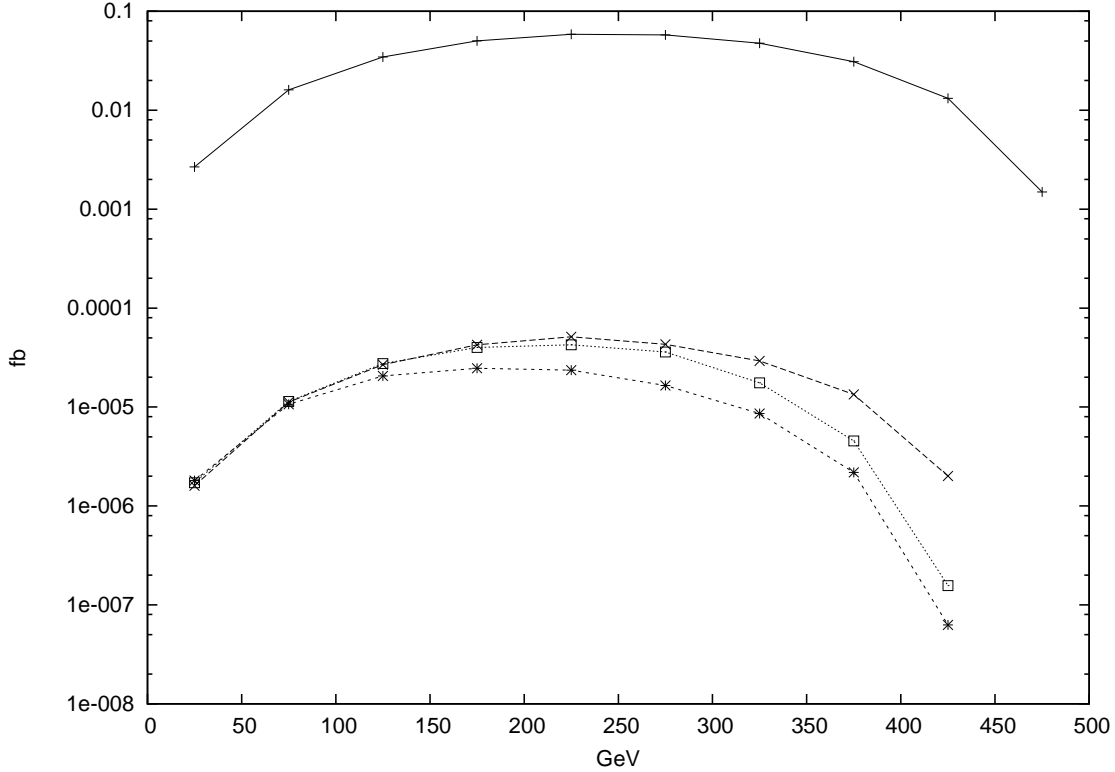}
\caption{The differential cross section $\Delta \sigma_{2 \Hp A}$ 
with respect to the charged Higgs momentum $q_{H^+}$. The horizontal
axis denotes $q_{\Hp}$ (GeV).
The long dashed line marked with cross symbol $\times$ corresponds to the case 
$(m_{H^+},m_A)=(200,200)$(GeV). The dotted
line marked with the boxes; $\Box$ 
corresponds to $(m_{H^+},m_A)=(300,200)$(GeV)
and the short dashed line marked by asterisks $*$ corresponds to
$(m_{H^+},m_A)=(200,300)$(GeV). The center of mass energy is $1000$
(GeV) and the width of each bin ($\Delta q_{\Hp}$) is $50$ (GeV).
For comparison, we also show the solid line with the plus sign $+$ for $W,Z$ pair production cross section,
$\Delta \sigma_{2 WZ}$ as a function of the momentum of W boson in final state $q_W$(GeV). For the cross section, the horizontal axis denotes the W boson 
momentum.}
\label{fig:sigma2}
\end{figure}
\begin{figure}
\includegraphics[width=15cm]{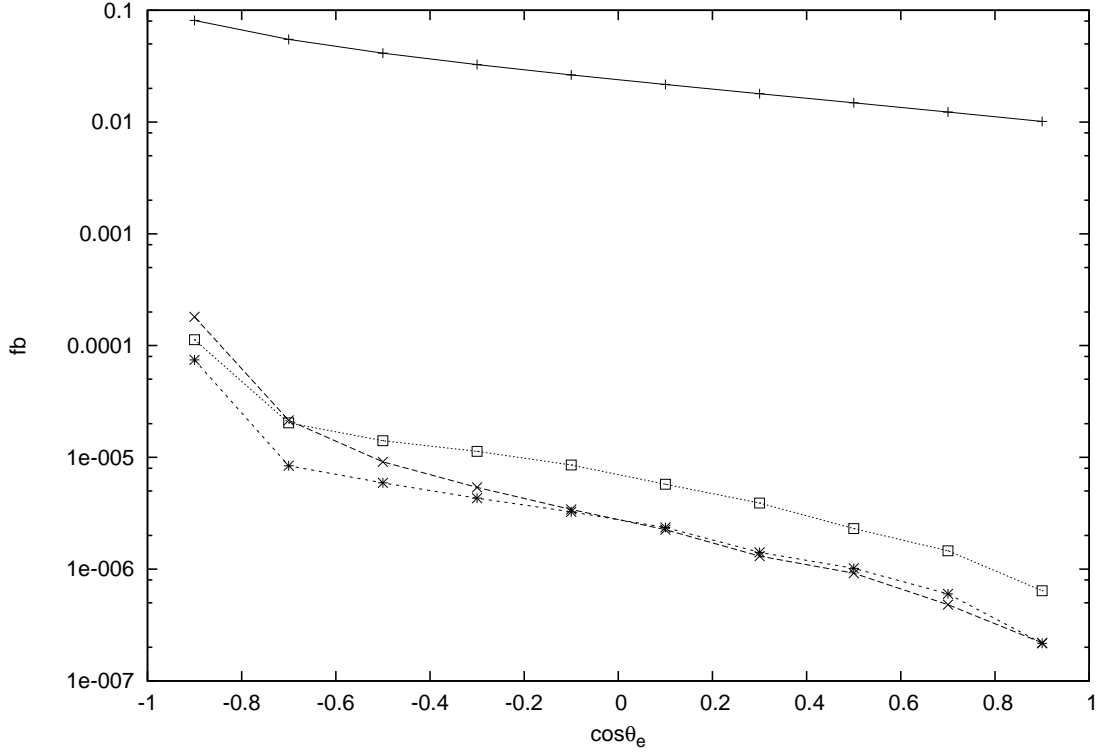}
\caption{The differential cross sections $\Delta \sigma_{3 \Hp A}$
for $e^+ + e^- \rightarrow H^+ + A + \overline{\nu_e}+ e^- $
with respect to
$\cos \theta_e $ where $\theta_e$ denotes the angle between
the final electron momentum and the initial positron momentum.
The long dashed line marked with cross symbol $\times$ corresponds to the case 
$(m_{H^+},m_A)=(200,200)$(GeV). The dotted
line marked with the boxes; $\Box$ 
corresponds to $(m_{H^+},m_A)=(300,200)$(GeV)
and the short dashed line marked by asterisks $*$ corresponds to
$(m_{H^+},m_A)=(200,300)$(GeV). The center of mass energy is $1000$
(GeV) and the width of each bin ($\Delta \cos \theta_e$) is 0.2.
For comparison, we show the cross section $\Delta \sigma_{3 WZ}$
of the process 
$e^+ + e^- \rightarrow W^+ + Z + \overline{\nu_e}+ e^-$ with solid line.
We use the formulae for the $W + Z \to W + Z$ scattering in \cite{Bahnik:1997ka}.
The center of mass energy of $e^+ e^-$ collision is $1000$(GeV).
}
\label{fig:sigma3}
\end{figure}
\begin{figure}
\includegraphics[width=15cm]{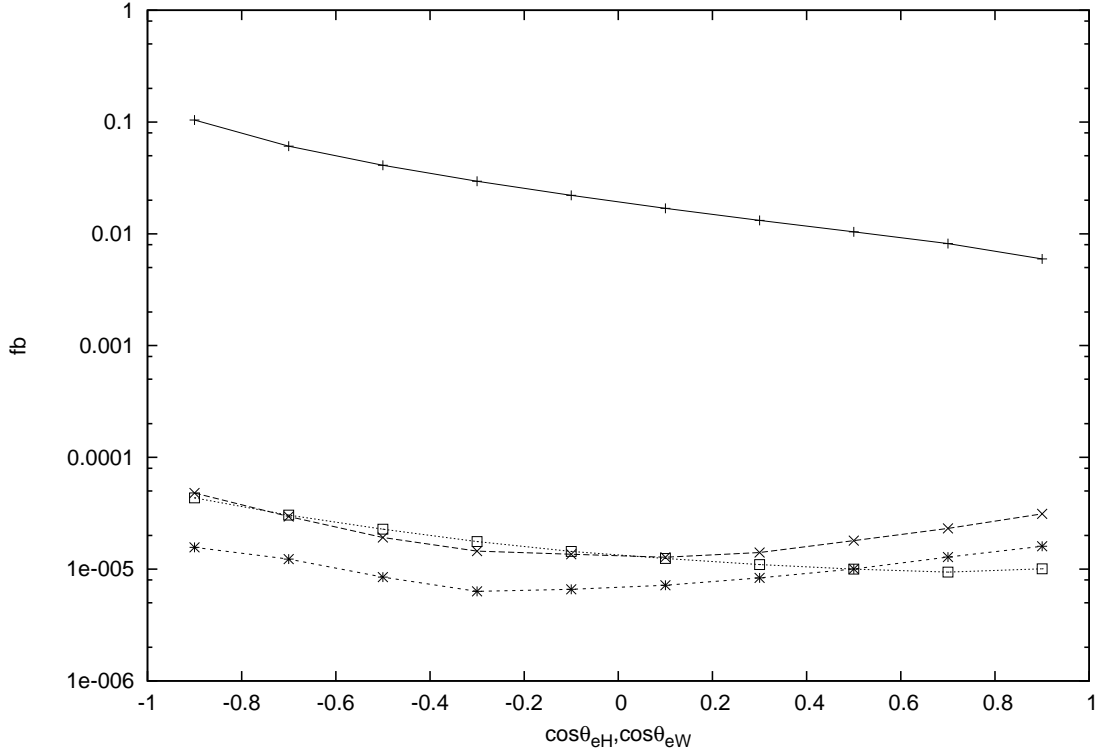}
\caption{Differential cross sections for 
$\Delta \sigma_{4\Hp A}$ and  $\Delta \sigma_{4 WZ}$. 
The horizontal axis corresponds to $\cos \theta_{eH}$ and $\cos \theta_{eW}$.
$\theta_{eH} (\theta_{eW})$ is an angle between the momentum 
of the final electron and the one of the charged Higgs boson
($W$ boson). 
The solid line marked with the plus sign $+$ corresponds to $W Z$ production.
The other three lines are Higgs pair production. Among them, 
the long dashed line marked with the cross symbol
$\times$ corresponds to the case 
$(m_{H^+},m_A)=(200,200)$(GeV). The dotted
line marked with the boxes; $\Box$ 
corresponds to $(m_{H^+},m_A)=(300,200)$(GeV)
and the short dashed line marked by asterisks $*$ corresponds to
$(m_{H^+},m_A)=(200,300)$(GeV). The center of mass energy is $1000$
(GeV) and the bin widths; $\Delta \cos \theta_{eH}$ and
$\Delta \cos \theta_{eW}$ are $0.2$.}
\label{fig:sigma4}
\end{figure}
\begin{figure}
\includegraphics[width=15cm]{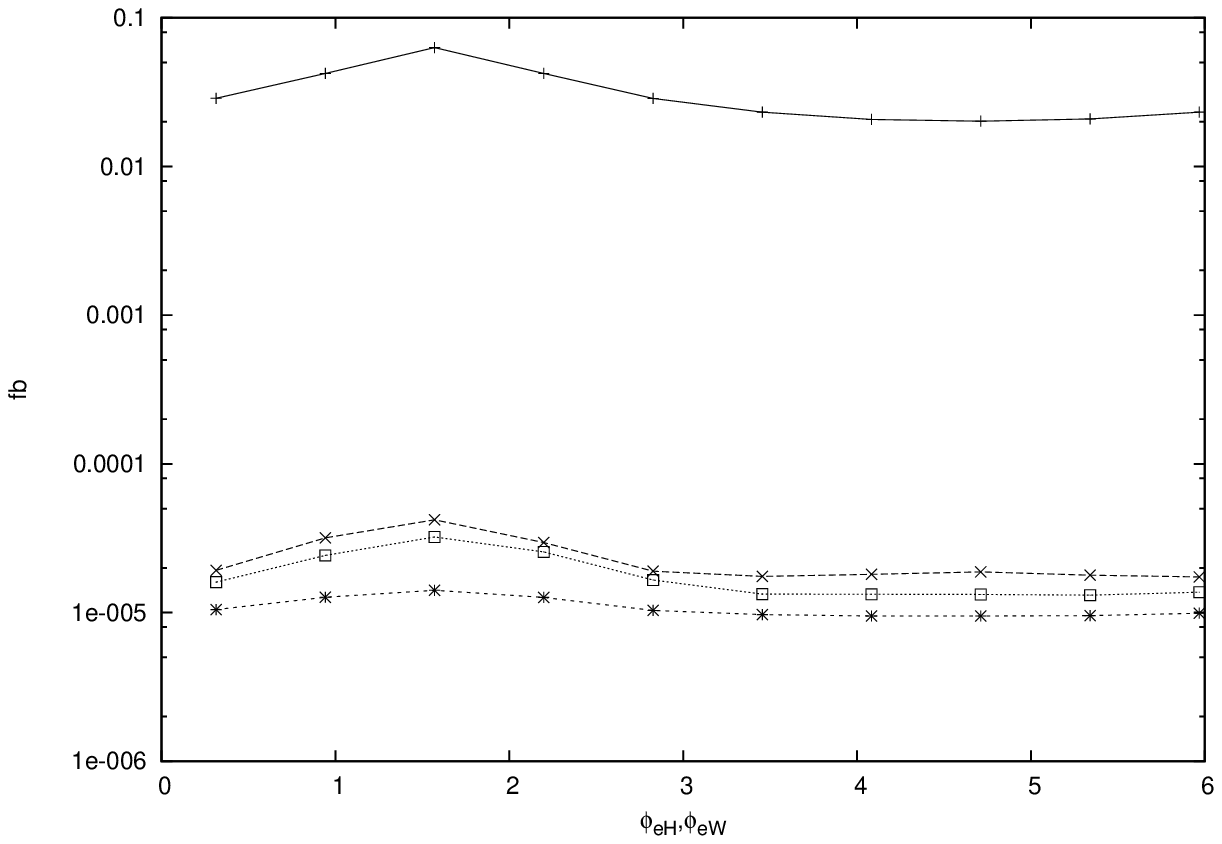}
\caption{Differential cross sections  
$\Delta \sigma_{5 \Hp A}$ and $\Delta \sigma_{5 WZ}$.
The horizontal line denotes the azimuthal angles $\phi_{eH}$ and $\phi_{eW}$
(radian). 
The solid line marked with the plus sign $+$ corresponds to $W Z$ production.
The other three lines are Higgs pair production. Among them, 
the long dashed line marked with cross symbol $\times$ corresponds to the case 
$(m_{H^+},m_A)=(200,200)$(GeV). The dotted
line marked with the boxes; $\Box$ 
corresponds to $(m_{H^+},m_A)=(300,200)$(GeV)
and the short dashed line marked by asterisks $*$ corresponds to
$(m_{H^+},m_A)=(200,300)$(GeV). The center of mass energy is $1000$
(GeV) and the bin widths; $\Delta \phi_{eH}$ and
$\Delta \phi_{eW}$ are $\frac{\pi}{5}$.}
\label{fig:sigma5}
\end{figure}

We summarize what one can read from figures of the cross sections.
(Fig.\ref{fig:cross} $\sim$ Fig.\ref{fig:sigma5}.) 
\begin{itemize}
\item The total cross section for Higgs pair production $\sigma_{\Hp A}$
increases as the center of mass energy of $e^+ e^-$ collision grows
until it reaches to 2000 (GeV). Even in the case for the lightest 
masses of Higgs pair which we have chosen, 
the cross section is at most $0.001$ fb. 
Compared with gauge boson pair production $\sigma_{WZ}$, the ratio
$\frac{\sigma_{\Hp A}}{\sigma_{WZ}}$ is order of $\sim 10^{-3}$.
\item The differential branching fraction with respect to the electron momentum in final states and with respect to the charged Higgs spectrum, 
they are limited
by phase space and for lighter Higgs pair masses, the momentum
of electron is larger. 
\item The distribution of the direction of the electron in the final states is 
strongly peaked at $\cos \theta_e=-1$. This implies the electron scattered 
into the forward direction with respect to the incoming electron.
This happens because virtuality of $Z^*$ boson is minimized in this case.
\item About the azimuthal $\phi_{eH}$ angle distributions, we find that
the charged Higgs momentum more likely lies within the range
$ 0 \le \phi_{eH} \le \pi$ than in $ \pi \le \phi_{eH} < 2 \pi$.  
\end{itemize}
\section{The signature of the charged Higgs and the neutral Higgs pair production}
\begin{figure}
\begin{center}
\includegraphics[width=10cm]{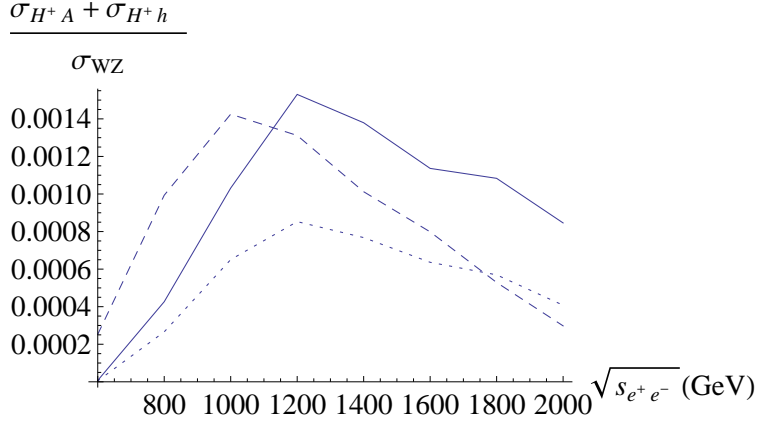}
\end{center}
\caption{The ratio of the cross sections
of Higgs pair production and gauge boson pair production;  
$\frac{\sigma_{H^+A}+\sigma_{H^+h}}{\sigma_{W^+Z}}$
as functions of center of mass energy of $e^+ e^- collision$;
$\sqrt{s_{e^+ e^-}}$(GeV). The solid line corresponds to the case
for $(m_H^+, m_A)=(300,200)$(GeV). The dashed line corresponds to
the degenerate case, $m_A=m_{H^+}=200$(GeV). The dotted line corresponds to the
case $(m_H^+, m_A)=(200,300)$(GeV).}
\label{fig12} 
%non-degenerate charged Higgs and neutral Higgs.
%We choose their masses as $M_{H^+}=300$(GeV), $M_{X}=200$(GeV) $X=h,A$.}
\end{figure}

\begin{figure}[htb]
\begin{center}
\includegraphics[width=10cm]{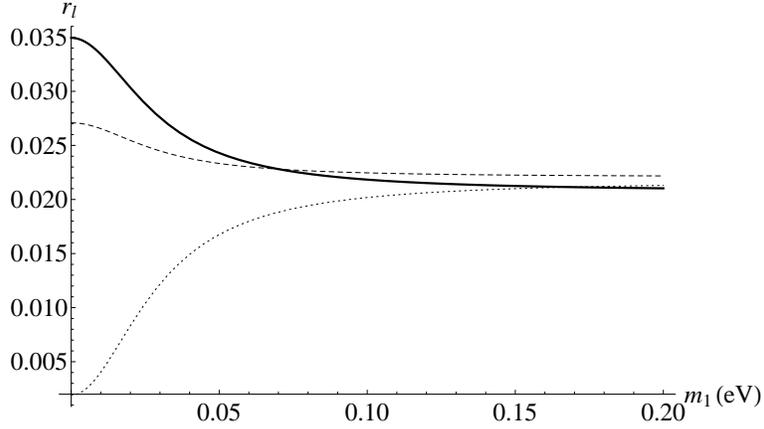}
\caption{ $r_l$ $(l=e, \mu, \tau)$ 
for normal hierarchical case as 
 functions of the lightest neutrino mass $m_1$
(eV). The dotted line corresponds to 
$r_e$, 
the dashed line corresponds to $r_\mu$ and 
the solid line corresponds to 
$r_\tau$ respectively.} 
\label{Fig13}
\end{center}
\end{figure}

As we have seen from the studies of the previous section,
the cross section and the differential cross sections of the Higgs pair
production are much smaller than gauge boson pair production.
Considering the smallness,
one may wonder if such Higgs pair production and their decays have 
the distinct signals.
Here we consider the charged lepton flavor dependence of the charged
Higgs decays into anti-lepton and 
neutrino. Note that the dominant neutral Higgses
decay channel is  neutrino and anti-neutrino pair when the neutral Higgs and
charged Higgs are degenerate as $|m_A-m_\Hp|< m_W$.
We study the degenerate case.
In this case, the neutral Higgs decay products are invisible and the visible
decay product is a charged anti-lepton $l^+$ from the charged Higgs decay.
Therefore, the whole process starting from $e^+ e^-$ collision to Higges decays
looks like,
\bea
e^+ + e^- &&\rightarrow \overline{\nu_e} + e^- + H^+ + A \nn \\
&&\rightarrow \overline{\nu_e} + e^- + l^+ \nu_l+
\nu_k \overline{\nu}_k.
\label{eq:signal}
\eea 
%In the gauge boson pair production process within the standard model,
One finds the same final state as in Eq.(\ref{eq:signal})
in gauge bosons pair production process of $e^+ e^-$ collision
as follows;
By replacing the charged Higgs boson with 
$W^+$ boson and the neutral Higgs boson $A$ with $Z$ boson in 
Eq.(\ref{eq:signal}),  
the decay channels $Z \rightarrow \nu_k \overline{\nu_k}$ and $W^+ \rightarrow l^+ \nu_l$ lead to the same final state as that of Eq.(\ref{eq:signal}).
\bea
e^+ + e^- &&\to 
\overline{\nu_e}+ e^- + W^{+}+ Z \nn \\
&&\to \overline{\nu_e} +e^- + l^+ \nu_l + \nu_k \overline{\nu_k}.
\label{eq:back}
\eea
Since Eq.(\ref{eq:back}) has a common final state with 
Eq.(\ref{eq:signal}), they look indistinguishable.
However 
as pointed in \cite{Davidson:2009ha}, the branching fraction of the
charged Higgs decay into anti-lepton is flavor non-universal and depends on the
lepton family. They are written in terms of the neutrino
mixings and masses which precise data except lightest neutrino mass
and CP violating phase is now available.
Since the W boson decay into anti-lepton is flavor blind,
we study the lepton flavor dependence of charged Higgs decay by taking 
the ratio with the weak gauge boson pair production and decay branching fractions.
The ratio we define is
\bea
r_l&=& \frac{\sum_{X=h,A}\sigma_{\Hp X} Br(X \to \nu \bar{\nu})
}{\sigma_{WZ}Br(Z \to \nu \bar{\nu})}
\frac{Br(H^+ \to l^+ \nu_l)}{Br(W^+ \to l^+ \nu_l)},
\label{eq:ratio}
\eea
where we used the shorthand notation,
$
Br(X \to \nu \bar{\nu})=\sum_{k}Br(X \to \nu_k \bar{\nu}_k),
$ 
for $X=h,A,Z$.
Using the notations, one can write $r_l$ as,
\bea
r_l=\frac{2 \sigma_{H^+ A}}{\sigma_{WZ}} \frac{Br(A \to \nu \bar{\nu})
}{Br(Z \to \nu \bar{\nu})}
\frac{Br(H^+ \to l^+ \nu_l)}{Br(W^+ \to l^+ \nu_l)},
\label{eq:rl}
\eea
where we use the fact that the production cross section for CP even and CP odd Higgs
with $U(1)$ charge is almost identical to each other; i.e.,
$\sigma_{H^+A} \simeq \sigma_{H^+h}$. (See appendix A). We also use the 
branching
fractions satisfy
\bea
Br(A \to \nu \overline{\nu})=Br(h \to \nu \bar{\nu})=100 \%.
\eea 
We have shown the ratio of cross sections in Fig.~\ref{fig12}.
When Higgs masses are degenerate $m_A=m_\Hp=200$ (GeV), the ratio of 
the cross section is about 
$1.4 \times 10^{-3}$ for $\sqrt{s_{e^+ e^-}}=1000$(GeV).
In what follows, we use this value as benchmark point
for the ratio of the cross sections in Eq.(\ref{eq:rl}).
The other branching fractions which appear
in Eq.(\ref{eq:rl}) are quoted from Particle Data Group (PDG) \cite{PDG:2012},
\bea
Br(W^+ \to \tau^+ \nu)&=&11.25 \pm 0.20 \%, \nn \\
Br(W^+ \to \mu^+ \nu)&=&10.57 \pm 0.15 \%, \nn \\
Br(W^+ \to e^+ \nu)&=&10.75 \pm 0.13 \%, \nn \\
%Br(Z \to b \bar{b})&=&15.12 \pm 0.05 \%, \nn \\
Br(Z \to \nu \bar{\nu})&=& 20.00 \pm 0.06 \%.
\eea
Using the numerical values, one can write $r_l (l=e, \mu,\tau) $ as,
\bea
r_e&=&0.465 \times
Br(H^+ \to e^+ \nu)\frac{2 \sigma_{H^+ A}}{
\sigma_{WZ}}
,\nn \\
r_\mu&=&0.473 \times
Br(H^+ \to \mu^+ \nu)\frac{2 \sigma_{H^+ A}}{
\sigma_{WZ}}
,\nn \\
r_\tau&=&0.444 \times
Br(H^+ \to \tau^+ \nu)\frac{2 \sigma_{H^+ A}}{
\sigma_{WZ}},
\eea
where $Br(H^+ \to l \nu)$ in $\%$ unit should be substituted.
The charged Higgs can decay into charged leptons and neutrino.
In contrast to the leptonic decay of W boson, the branching fractions 
for each flavor of charged lepton are obtained from Eq.(\ref{eq:Yukawa})
\cite{Davidson:2009ha},
\bea
Br(H^+ \to l^+ \nu_l)=\frac{\sum_{i=1}^3 {m_i^2 } |V_{li}|^2}{\sum_{i=1}^3 
m_i^2} \times 100 \%.
\eea
We update the branching fraction to each lepton flavor mode
using the recent results on $|V_{e3}|$.
For normal hierarchy case, the branching fractions are written as,
\bea
{\rm Br}(H^+ \to l^+ \nu_l)=\frac{ m_1^2+\Delta m_{sol.}^2 |V_{l2}|^2
+(\Delta m_{sol}^2 + \Delta m_{atm}^2) |V_{l3}|^2}{3 m_1^2
+ 2\Delta m_{sol}^2 + \Delta m_{atm}^2}\times 100 \%.
\label{eq:normal}
\eea
In the formulae of Eq.(\ref{eq:normal}), $m_1$ denotes the lightest neutrino mass.
For inverted hierarchical case, they are written as,
\bea
{\rm Br}(H^+ \to l^+ \nu_l)=\frac{ m_3^2+\Delta m_{atm.}^2 
(|V_{l1}|^2+|V_{l2}|^2)
-\Delta m_{sol}^2 |V_{l1}|^2}{3 m_3^2
+ 2\Delta m_{atm}^2 - \Delta m_{sol}^2} \times 100 \%,
\eea
where $m_3$ denotes the lightest neutrino mass.
We have used the following values for the mixing angles and mass squared 
differences quoted from 
Table 13.7 in the review section of Neutrino Mass, Mixing, and Oscillation
of \cite{PDG:2012},
$
\sin^2 \theta_{12}=0.306, \ 
\sin^2 \theta_{23}=0.42, \ 
\sin^2 \theta_{13}=0.021,$
$m_{atm}^2=2.35 \times 10^{-3}$(eV$^2$) and 
$m_{sol}^2=7.58 \times 10^{-5}$ (eV$^2$).
In Fig.~\ref{Fig13}, we have shown $r_l$ ($l=e, \mu, \tau)$ for normal hierarchical case as functions of
the lightest neutrino mass $m_1$. 
\begin{figure}[htb]
\begin{center}
\includegraphics[width=10cm]{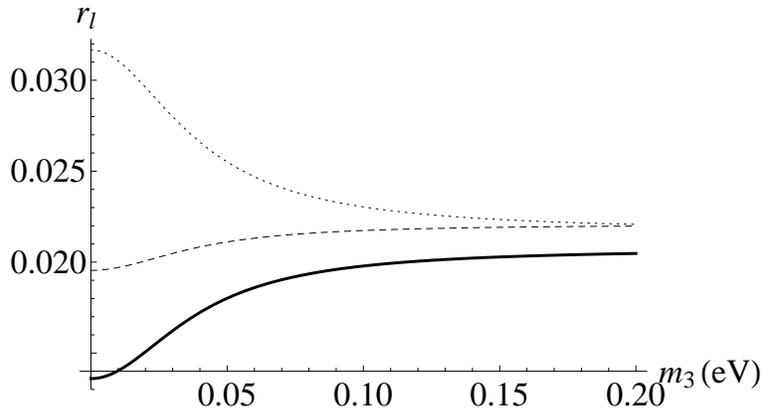}
\caption{ $r_l$ $(l=e, \mu, \tau)$ 
for inverted hierarchical case as 
 functions of the lightest neutrino mass $m_3$
(eV). The dotted line corresponds to 
$r_e$, 
the dashed line corresponds to $r_\mu$ and 
the solid line corresponds to 
$r_\tau$ respectively.}
\label{Fig14}
\end{center}
\end{figure}
In Fig.~\ref{Fig14}, we have shown $r_l$ for inverted hierarchical
case as functions for the lightest neutrino mass $m_3$. 
As we have seen from Fig.~\ref{Fig13} and Fig.~\ref{Fig14}, we can expect 
 $2 \% \sim  3 \%$ lepton flavor dependence from charged
Higgs decay. We summarize the flavor dependence.
\begin{itemize}
\item For normal hierarchical case, for $0 \le m_1 <0.05$(eV)
$r_\tau > r_{\mu}>> r_e$. For larger $m_1$ until $0.2$ eV,
$r_\mu \sim  r_e \sim r_\tau=0.02$.
\item For inverted hierarchical case,
$r_e >r_\mu> r_\tau$ for $0<m_3<0.2$ eV. 
\end{itemize}
\section{Conclusions and Discussions}
In this paper, we study the pair production of
charged Higgs and neutral Higgs in the neutrinophilic 
two Higgs doublet model. 
The pair production process is not suppressed by the U(1) charge
conservation. In other words, the approximate global symmetry
allows the pair production to occur.

We study the total cross section for the pair production 
in $e^+ e^-$ collision. The pair production occurs through 
W boson and Z boson fusion.
We study the pair production and the decays
for degenerate mass of charged Higgs and neutral Higgs
as well as non-degenerate case. The cross section increases from
 $10^{-4}$ fb to $10^{-3}$ fb as 
the cm energy of $e^+ e^-$ varies from 1 (TeV) to 2 (TeV).
The cross section is compared with that of $W,Z$ pair production.
We show the Higgs pair production is 
about $10^{-3}$ times smaller than the pair production cross section of the
gauge bosons. Therefore if Z decays invisibly into neutrino pairs and
W boson decays into anti-lepton and neutrino, the gauge boson 
pair production and their decays becomes a background to the signal.
When the charged Higgs ($H^+$)and neutral Higgs ($X=A, h$)
are degenerate as
$|m_{H^+}-m_X|< M_W$, which is favored from the electroweak precision
data,  the charged Higgs dominantly decays into anti-lepton and neutrino
and neutral Higgs decays dominantly into neutrino and anti-neutrino pair.
Compared with them, W and Z decay branching ratio in the same
final state is smaller than that of Higgs decays and is flavor blind.
Therefore, by studying  
the charged anti-lepton flavor in the final state, 
we may distinguish the Higgs pair production and their decays from that of
gauge bosons.
  We expect $2 \% \sim 3 \%$ flavor dependence
which is null for the gauge bosons decays. 
Depending on the normal or inverted hierarchy of the mass spectrum
of neutrinos, the order of $r_e,r_\mu$ and $r_\tau$ changes.
We show the differential cross sections with respect to electron, charged Higgs momentum.
The differential cross sections with respect to
the angles of electron 
and charged Higgs in the final states are also shown. 
They are also important to identify the signals.

\appendix
\section{Amplitude for $W^{+\ast} + Z^{\ast} \to H^+ + h$}
In this appendix, we have shown the off-shell charged Higgs
and CP even neutral Higgs (h) boson production amplitude
for gauge boson fusions
$W^{+\ast} + Z^\ast \to H^+ + h$.
\bea
T_{h \mu \nu}&=& \frac{g^2 \cos(\beta+\gamma)}{2\cos \theta_W}
\left(a_h g_{\mu \nu}+ d_h q_{h \nu} q_{H^+ \mu}
+b_h  q_{H^+ \nu} q_{h \mu}\right),
\label{eq:hamp}
\eea 
where we compute the four Feynman diagrams corresponding to , the contact interaction (Fig.~2), the s channel $W^+$
exchange(Fig.~3)
u channel charged Higgs exchange (Fig.~4), and 
t channel CP odd Higgs (A) exchanged diagram (Fig.~5).
$a_h, b_h$ and $d_h$ in Eq.(\ref{eq:hamp}) are given as,
\bea
a_h&=&-s_W^2-\frac{p_Z^2-p_W^2}{M_z^2}\frac{M_h^2-M_{H^+}^2-M_W^2}{s_{\Hp h}-M_W^2}-c_W^2
\frac{t_h-u_h+p_Z^2-p_W^2}{s_{\Hp h}-M_W^2}, \nn \\
b_h&=&\frac{2\cos 2 \theta_W}{u_h-M_{H^+}^2}+ 
\frac{2(\cos 2 \theta_W+1)}{s_{\Hp h}-M_W^2},\nn \\
d_h&=&-\frac{2}{t_h-M_A^2}-\frac{2(\cos 2 \theta_W+1)}{s_{\Hp h}-M_W^2},
\eea
with $t_h=(q_\Hp-p_W)^2, u_h=(p_W-q_h)^2$ and $s_{H^+h}=(q_\Hp+q_h)^2$.
By taking the vanishing limit of the U(1) breaking term; i.e., $m_{12} \rightarrow 0$, $\beta$ and  $\gamma$ vanishes. Note also in this limit, one can show
$m_h=m_A$ and $-iT_{A\mu \nu}=T_{h \mu \nu}$ with the appropriate replacement
$q_A \rightarrow q_h$. (See Eq.(\ref{eq:Aamp}).) Therefore in this limit
the production amplitudes for $H^+ A$ and $H^+ h$ are 
identical to each other,
$\sigma_{\Hp A}=\sigma_{\Hp h}$.
\section*{Acknowledgment}
%\noindent{\bf Acknowledgement}
We would like to thank H. Umeeda for discussion.
We also would like to thank M. Okawa and K. Ishikawa 
for their help on the numerical computation.
T. M. was supported by KAKENHI, Grant-in-Aid for 
Scientific Research(C) No.22540283 from JSPS, Japan.
\vfill\pagebreak

\end{document}